# Galactic forcing increases origination of marine microplankton

**Péter Ozsvárt[1]\*, Emma Kun[2,3,4,5], Imre Bartos[6], Zsolt G Márka[7,9], Szabolcs Márka[8]**

[1] HUN-REN-MTM-ELTE, Research Group for Paleontology, PO BOX 137, 1431 Budapest, Hungary, ozsvart.peter@nhmus.hu.

[2] Astronomical Institute, Faculty for Physics and Astronomy, Ruhr University Bochum, 44801 Bochum, Germany, ekun@astro.ruhr-uni-bochum.de

[3] Theoretical Physics IV: Plasma-Astroparticle Physics, Faculty for Physics and Astronomy, Ruhr University Bochum, 44801 Bochum, Germany

[4] Konkoly Observatory, HUN-REN Research Centre for Astronomy and Earth Sciences, H- 1121 Budapest, Konkoly Thege Miklós út 15-17., Hungary

[5] CSFK, MTA Centre of Excellence, Konkoly Thege Miklós út 15-17., Hungary

[6] Department of Physics, University of Florida, PO Box 118440, Gainesville, FL 32611-8440, USA

[7] Department of Earth Sciences, University of Cambridge, Downing Street, Cambridge CB2 3EQ, United Kingdom

[8] Department of Physics, Columbia University in the City of New York, New York, NY, USA

[9] Department of Earth and Planetary Sciences, University of California, Berkeley, CA, USA

Corresponding author: Péter Ozsvárt (ozsvart.peter@nhmus.hu)

**Key Points:**

- Significant correlation between the Solar System's oscillation and quasi-periodic blooming of marine microorganisms throughout Phanerozoic

- Cosmic ray air showers may potentially trigger DNA mutations in marine microplanktons resulting in origination of new genera/species

- Galactic forcing significantly influenced marine biodiversity through quasi-periodic cosmic ray exposure over the Phanerozoic.

**Abstract**

The continuous flux of Galactic cosmic rays that bombard Earth's atmosphere creates ionizing radiation that can damage the DNA of living organisms. While this radiation on Earth is relatively constant in the short term, large and long-scale fluctuations are expected with a period of ~63.5 million years. As the Solar System moves above or below the Galactic plane during its oscillatory motion about the Galactic center, the Galactic magnetic shielding weakens, allowing more cosmic rays to reach Earth and trigger mutations in organisms. We identify a significant correlation (weighted global p-value: $1.25 \times 10^{-4}$, or $3.72\sigma$) between the Solar System's Galactic oscillations and the origination of marine zoo- and phyto-microplankton genera over the Phanerozoic. When we restrict the analysis to time intervals during which all four groups coexisted, a post-trial significance of $4.52\sigma$ emerges. Our findings suggest that changes in biodiversity have been significantly influenced by long-term Galactic forcing.

**Plain Language Summary**

We observe a significant correlation between variations in cosmic ray flux due to the Solar System's oscillation above/below the Galactic plane and the quasi-periodic blooming of marine microorganisms throughout the Phanerozoic.

## 1. Introduction

Biodiversity on Earth has regularly undergone dramatic changes over the past ~550 million years (e. g. Raup & Sepkoski, 1982; Rohde & Muller, 2005; Roberts & Mannion, 2019). Fluctuations in biodiversity over the Phanerozoic are characterized by five major mass extinction events in the paleontological record (Raup & Sepkoski, 1982), however, the Phanerozoic has also seen large-scale diversification of the biosphere (Rohde & Muller, 2005). Some of these changes were caused by temporally localized events – such as asteroid impacts (Raup & Sepkoski, 1982 volcanic activity (Hernandez Nava et al., 2021*)*, or dramatic cosmic phenomena (Svensmark, 2023; Nojiri et al., 2025) – while others may have been due to long-scale cyclic events such as plate fragmentation (Zaffos et al., 2017) or sea-level change (Roberts & Mannion, 2019; Boulila et al., 2023). A number of studies have observed strong cyclicities in fossil data (Rohde & Muller, 2005; Roberts & Mannion, 2019; Boulila et al., 2023*,* Melott & Bambach, 2013), though the presence of many of these cyclicities – especially between extinction events – have been debated (e.g., Bailer-Jones, 2009). In addition to the aforementioned causes (Zaffos et al., 2017*,* Roberts & Mannion, 2019; Boulila et al., 2023) for long-term periodicities in the paleontological record, astronomical causes (e.g., cosmic rays, varied comet or meteorite flux, and supernovae) have also been proposed in the literature (Svensmark, 2023; Nojiri et al., 2025). Cosmic phenomena are one of the prime candidates for the source of long-term fluctuations in biodiversity, as their effects on Earth are modulated by the passage of the Solar system through the Milky Way's plane or its spiral arms (Bailer-Jones, 2009; Alroy, 2008; Bambach, 2006). Although cyclicities in extinction events over the Phanerozoic, and their connection to cosmic phenomena have been largely disproven by multiple studies (Svensmark, 2023; Nojiri et al., 2025), there are many other possible expressions of galactic forcing of the biosphere, namely species origination (Mayhew et al., 2008), changing diversity (Hannisdal & Peters, 2011), blooming events, or the accelerated evolution of different taxa (De Wever et al., 2001).

Here, we observe the effects of, and provide a plausible mechanism for, the galactic forcing of Earth's biosphere by analyzing datasets of marine microplankton fossil occurrences over the Phanerozoic for cosmic ray-induced cyclicity due to the Solar System's movement above/below the galactic plane with a period of ~63.5 Ma (Fig. 1).

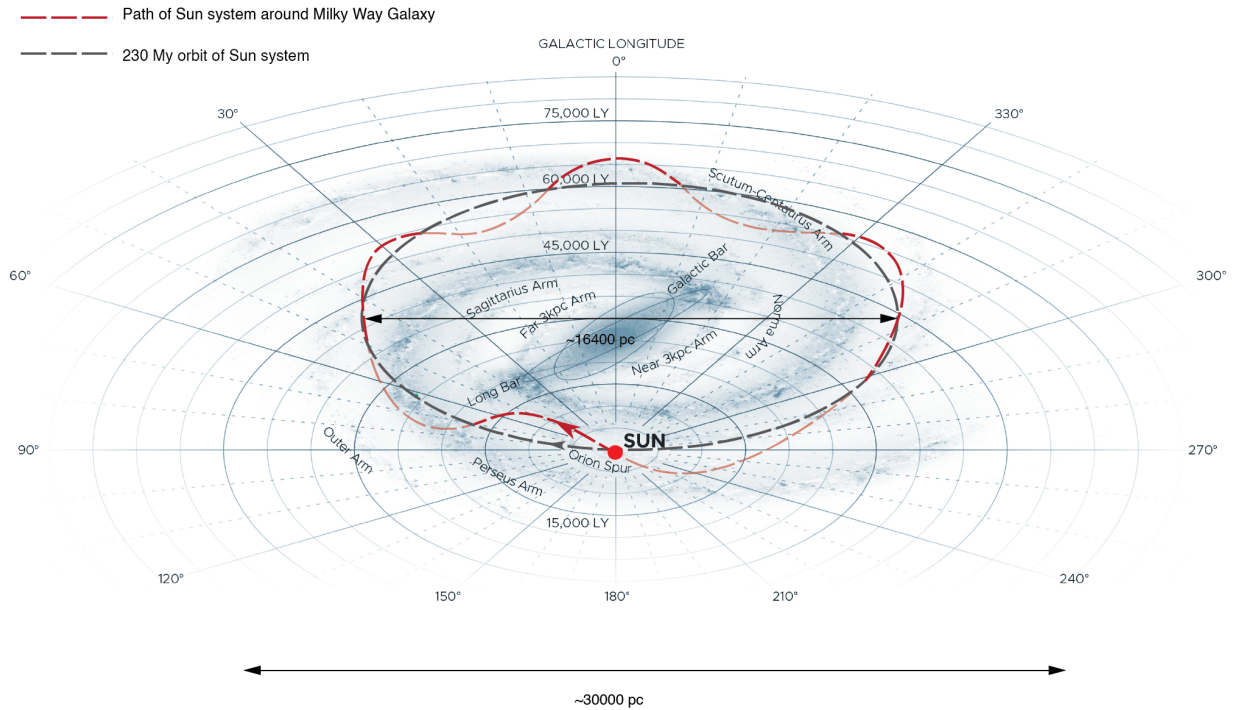

Figure 1. Solar system moving in the Galaxy as viewed from above the plane (Original credit: NASA/JPL-Caltech/R. Hurt (SSC/Caltech).

Marine microplankton (phyto- and zooplanktons) play a significant role in photosynthesis, organic carbon production, and are the main suppliers of the carbon cycle in marine ecosystems throughout the Phanerozoic. These marine planktonic organisms have complex, unique, and diverse fossilizable carbonate (planktonic foraminifera or nannoplankton), siliceous (radiolaria) or organic (dinoflagellate) skeletons, and are widely distributed from the warm surface tropical waters to the cold seas of the high latitudes. Furthermore, the current state of knowledge on these groups have recently been compiled by multiple groups and published in well-curated and frequently updated databases (see Open Research and Caridroit et al., 2017; O'Dogherty et al., 2009; 2021), from which the first stratigraphic appearance and ranges of individual genera/ species can be extracted with stage or sub-substage precision, allowing us to analyze fossil data for periodicities with greater certainty than before (Fig. 2). Of the four groups examined, the evolutionary history of radiolarians is representative of the entire Phanerozoic while nannoplankton, dinoflagellates and planktonic foraminifers have been present in the marine ecosystem for nearly 200 million years. In addition to the new databases, these microplanktonic groups were chosen as the target of our study, as, due to their skeleton size (2 μm - 300 μm) and their mass distribution in near-surface waters, they are more exposed to the effects of various cosmic phenomena (e.g. cosmic ray forcing effect) and their evolutionary history is well-documented.

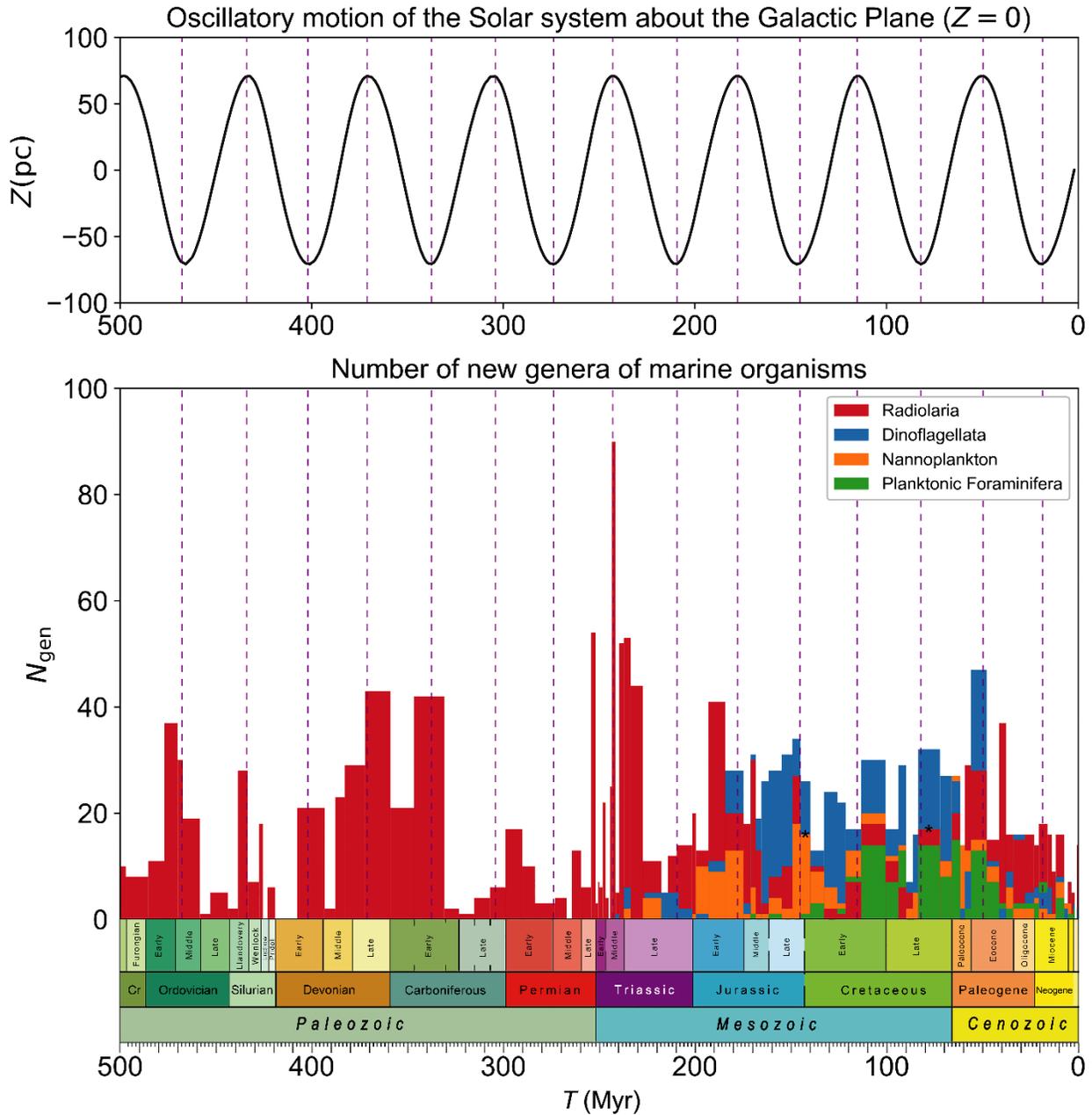

Figure 2. Origination for Phanerozoic Radiolaria (red), Dinoflagellata (blue), Nannoplankton (orange) and Planktonic Foraminifera (green) genera compared to the oscillatory motion of the Solar system through the plane of the galaxy. The width of the bars in the histogram represent the duration of each Stage/ Substage. Z is the distance from the Galaxy plane in parsec (pc). Chronostratigraphy from the GTS 2020 (Gradstein et al., 2020). * = equal number of new radiolarian and nannoplankton genera originating in the Berriasian and Campanian stages. The Paleozoic radiolarian database includes species marked as "nomina dubia" from the Aitchison et al. 2017 database.

## 2. Materials and Methods

### 2.1 Choice of databases and geological time scale

To construct our origination curves, we assembled genera data from databases that are regularly updated with the latest published data, and where taxonomic assignments are regularly reconsidered (see below). By doing so, we aimed to make use of the most complete and up-to-date datasets whilst also minimizing the need for data filtering. All these catalogs and databases have been created by immense efforts to represent the current state of the taxonomy and stratigraphic ranges of all known genera by the consensus of multiple active working teams.

It is for this reason that we opted to use these databases instead of ones such as the PDBD, Neptune, or Sepkoski's compendium that, while fantastic tools for biodiversity studies due to their large data volumes, rely on authors to submit new data and do not have regular updates for uncertain taxonomic or formation/stratigraphic assignments. Though the use of these continually updated datasets minimizes errors coming from incorrect taxonomic, stratigraphic, or age assignments, they, like any fossil database, are subject to significant biases caused by varied spatial or temporal fossil preservation (Smith & McGowan, 2007). As we do not have access to complete sample size information, we were unable to apply sampling standardization to the datasets, thus the direct measurement of true origination is uncertain. The origination histogram reconstruction was obtained from the first appearance datum (FAD) of the different databases (Fig. 2). We aimed to produce the highest resolution origination histogram by extracting genera at the smallest possible level (Stage/Substage) from the below databases. This means that the longest time epoch over which the first appearance datums are 'binned' is the Early Cambrian Terreneuvian Epoch (17.80 Ma), and the shortest is the Early Triassic Dienerian (0.6 Ma). On average, we have an origination record of four different microplanktonic groups every 5.79 million years. The age assignments of genera originations are, of course, another significant potential source of error, with previous examinations of fossil genera originations from the Sepkoski compendium using the older ICS 2004 timescale not showing ~63Ma cyclicities when calibrated to more recent timescales (Rohde & Muller, 2005; Alroy, 2008). We, therefore, calibrated the obtained data to the most recent geologic timescale from Gradstein (2020). It is important to note that evolutionary trends in the paleontological record are likely biased by variable fossil preservation, changes in the abundance of fossil-forming rocks over time, or inaccurate stratigraphic correlations. When analyzing large fossil datasets for long-scale periodicities it is, therefore, important to account for such preservation biases when attempting to make causal inferences for the source of macroevolutionary trends (Kidwell & Holland, 2002).

### 2.2 Methods for hypothesis tests

To test the connection between the bloomings of the four marine groups and the time of maximum excursions of the Solar system from the Galactic plane, we carried out a one-sided Monte Carlo permutation test of the hypothesis that new genus appearances are associated with the maximum excursion of the Solar System from the Galactic plane. This decision was guided by the expectation that the emergence of new genera may not follow a smooth, sinusoidal pattern, due to the expected shape of the cosmic ray flux variations, but rather exhibit asymmetric behavior, characterized by sharp peaks (e.g., diversification events) and broad troughs (e.g., relative stasis). The null hypothesis states that there is no real association between the appearance of new genera and the Solar System's position relative to the Galactic plane. The test-statistic was the sum of new genera appearing in each geological stage/ substage, during which the Solar system was maximally off to north or south from the Galactic plane ($\sum N_{gen,obs}$). We constructed the (discrete) probability density function (PDF) of $\sum N_{gen,rnd}$ by randomly assigning the number of new genera to the geological ages $10^5$ times. With this we made a resampling under the null hypothesis. Then integrating the PDF from the observed value of $\sum N_{gen,obs}$ to positive infinity, we obtained the tail-probability of observing $\sum N_{gen,rnd}$ as extreme or even more extreme than the observed one ($\sum N_{gen,obs}$). To calculate a global p-value, we combined the four p-values employing the Stouffer-method (Stouffer et al., 1949; Zaykin, 2011). The Stouffer method uses the weighted average of the individual tests' z-score values, and it can be applied when the p-values are correlated. We report two global p-values, one that makes use of the whole dataset, and one that assesses the time period when all four groups coexisted.

## 3. Cosmic ray driver

Although the precise mechanisms and causes of rapid microplankton origination events over the Phanerozoic are not yet fully clear, our study reveals a more complex pattern of biotic and abiotic dynamics than has previously been appreciated. Our model suggests that one of the most direct external drivers of origination might be an increase in the exposure of the biosphere to cosmic radiation (Medvedev & Melott, 2007). Cosmic radiation is produced by electrically charged, ionizing particles, which originate from either our Solar System (mainly from the Sun, at keV, MeV energies) or Galactic and extragalactic sources (MeV–EeV energies). The role of cosmic rays is important in galactic dynamics as they heat interstellar gas, influence star formation, and since charged cosmic rays interact with magnetic fields. The observed energy spectrum of cosmic rays spans about 12 orders of magnitude from 0.1 GeV to slightly more than $10^{11}$ GeV, and the flux axis ranges from 4 particles/cm²/s to 1 particle/km²/100 years (Pierre Auger Collaboration, Fenu, 2017). According to composition measurements, the spectrum is strongly proton-dominated up to ~100 GeV. There is, therefore, a high level of heterogeneity in cosmic ray distribution and energy as Solar system, Galactic, and extragalactic sources mix.

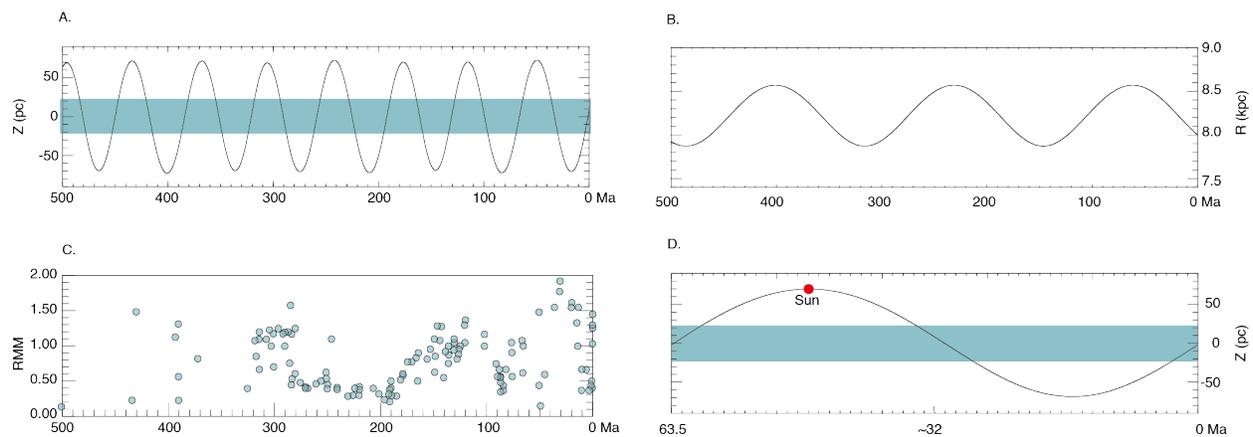

Figure 3. The Sun system's oscillatory motion in our galaxy. A: Solar system's oscillatory motion through the plane of the galaxy in the past 500 Ma (from Gies & Helsel, 2005); B: Solar system's distance from the Galactic center (from Gies & Helsel, 2005); C: Intensity of the magnetic field in the past 500 Ma (from Denis et al., 2002); D: The period of the Solar system oscillation perpendicular to the Galactic disk $P_z$ =

63.5 Ma and the maximum oscillation of the Solar system from the Galactic plane (from Gies & Helsel, 2005), where the magnetic shielding of cosmic rays is expected to be strongest.

Our Solar system orbits around the Milky Way galactic center with a period of about 230 million years (Fig. 1) (Bland-Hawthorn & Gerhard, 2016). The Solar System is not just simply in a flat orbit; it oscillates vertically and radially in the Galactic disk (Fig. 3A-B) with a period of about 63.5 million years and with a vertical amplitude in the order of ~70 pc (Atri & Melott, 2011). The Solar system oscillates through distinct regions of the thin and inhomogeneous (Bland-Hawthorn & Gerhard, 2016) Galactic disk (composed dominantly of gas, dust or young stars), where a large flux of Galactic cosmic rays with energies up to $10^{15}$ eV is expected (Dörner et al., 2024) (Fig. 1). When the Solar System is located well within the Galactic plane, the predominantly toroidal galactic magnetic field protects the Solar System from cosmic radiation (Oppermann et al., 2012). Though the Galactic magnetic field is weak (a few μG e. g. Pandhi et al., 2022), this effect is large enough to limit cosmic ray exposure. Given the heterogeneous nature of the thin Galactic disk (Oppermann et al., 2012), the Solar system's oscillation north or south out of the Galactic disk plane (Fig. 1 and 3D), exposes it to more-and-more high-energy cosmic rays due to (a) the heterogeneous composition of the thin disk, and, more importantly, (b) decreased Galactic magnetic shielding as it deviates from the core of the Galactic disk. Assuming the scaling factor for cosmic ray flux attenuation is $\kappa_0 = 0.05$ μG$^{-2}$, the scale height as d=1 kpc, and the magnetic field strength as $B_0 = 6$ μG (Beck, 2009), we calculate that the cosmic-ray flux at 1GeV ($10^5$ GeV) drops by about 15% (1%) when the Solar system is at its maximum oscillation as compared to the cosmic-ray flux in the Galactic plane. Though the bulk of the particles reaching Earth are deflected back into outer space by the heliosphere and Earth's magnetosphere, part of their flux is able to impinge the upper atmosphere, leading to a shower of secondary particles that reach Earth's surface (The Pierre Auger Cosmic Ray Observatory, 2015; Abu-Zayyad et al., 2012). The muon component dominates in these secondary particles, contributing 85% of the radiation dose from cosmic rays (Atri & Melott, 2011). The radiation dose can also be modulated by ozone depletion, which increases the ultraviolet flux reaching Earth's surface (Melott & Thomas, 2011). Simulations indicate that proton showers with energies of $10^4$-$10^5$ GeV – energies at which protons dominate the observed galactic cosmic-ray spectrum (Schroeder et al., 2019) – lose their energy within the first ~10 meters of seawater (Sloan, 2007), where one expects to find the bulk of marine planktonic microorganisms (De Wever et al., 2001).

Cosmic rays do not directly interact much with water. Instead, they initiate air showers when they enter the Earth's atmosphere, creating a cascade of secondary particles (such as muons, electrons, neutrinos, and gamma rays), which do interact with water. Neutrinos, produced in cosmic ray interactions, rarely interact, but when they do in water, they can create charged leptons (like muons or electrons), only subdominantly contributing to the electron flux anywhere in water. High-energy cosmic ray secondaries (especially neutrons or pions) can cause spallation reactions in the oxygen nuclei in water, creating radioactive isotopes. We see that cosmic rays and, more dominantly, air showers initiated by cosmic rays, can interact with water, leading to an excess of solvated electrons (e.g., Kumar et al., 2019) that can cause frequent gene mutations and thereby more genera originations.

### 3. 1 Statistical probe of microplankton originations

To quantify the effect of variable cosmic ray flux on the origination of dinoflagellates, nannoplankton, planktonic foraminifera, and radiolarians, we carried out one-sided Monte Carlo permutation tests of the hypothesis that new genus appearances are associated with the maximum excursion of the Solar System from the Galactic plane (see Material and Methods). Four different independent microplanktonic groups were analyzed, two of which were phytoplankton (nannoplankton and dinoflagellate) and two zooplanktons (radiolaria and plankton foraminifera). Estimating these origination curves during the past 500 million years is a serious challenge due to the extreme heterogeneity of published data (Foote, 2000). Recently available databases have enabled us to analyze more comprehensive origination time series than were previously available (Fig. 2). However, they contain only limited information for estimating origination over the entire Phanerozoic. Only radiolarians provide continuous datasets from the Early Cambrian to the present, while the other different microplanktonic groups only provide plausible data from the middle Mesozoic, as these groups appeared much later (see Material and Methods). The origination record exhibits several steps and peaks that reflect short episodes of appearances of a variable number of new genera. Rapid blooms and accelerated evolutionary trends characterize the evolutionary history of radiolarians during the entire Phanerozoic, which is generally followed by a long period of evolutionary equilibrium. We also see a consistent trend for other microplankton groups from the middle Mesozoic Era. As the exact period of the Solar system's oscillation north and south of the Galactic plane is not known, we repeated the analysis while introducing systematic shifts of -5 to +5Ma to the oscillation period.

## 4. Results

The timing of the appearance of new nannoplankton, dinoflagellate, radiolaria, and planktonic foraminifera genera follow that of the oscillation of the Solar system above/below the Galactic plane (see Fig. 2), and therefore the weakest magnetic shielding of the biosphere from cosmic rays (see Fig. 2). The *p*-values resulting from the statistical probe executed (see Materials and Methods) indicate that the observed blooming events are highly unlikely to be randomly associated with maximal Solar system oscillation from the Galactic plane for dinoflagellates, nannoplankton and planktonic foraminifera (see Fig. 4a).

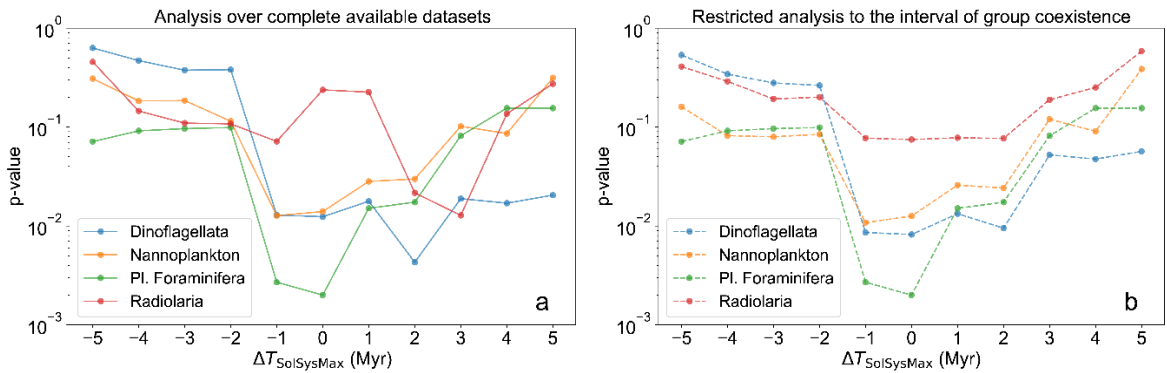

Figure 4. Random chance of finding a high number of new genera in four marine groups (Dinoflagellata, Nannoplankton, Planktonic Foraminifera and Radiolaria) during maximum excursions of the Solar system from the Galactic plane. We plot the p-values as a function of the shift introduced at the time of maximum Solar system excursions. a: Results for the whole databases, b: results, when we cut the length of the databases at the appearance time of the planktonic foraminifera.

To assess whether the groups are all affected by the same concerted periodic effect, we repeated the statistical analysis while introducing ad-hoc shifts of -5 to +5Ma to the initial phase of the oscillatory motion of the Solar system. The trend in *p*-values when adding these shifts to the Solar system oscillation peaks is similar for dinoflagellates, nannoplankton, and planktonic foraminifera, indicating that these groups are affected by the same periodic effect. The radiolarian *p*-value trend differs, largely due to the additional ~300Ma of origination data (and the consequent greater uncertainties in age constraints, preservation bias, etc.) relative to the other groups. To confirm the common forcing mechanism influences radiolarian occurrences as well, we repeated the full analysis in only the time periods when all four groups were present concurrently (i.e., the length of shortest database, planktonic foraminifera, 0-174 Ma) and observed similar trends in significance, indicating a shared driving mechanism (see Fig. 4b). Another factor impacting radiolarian *p*-value trends over ~500Ma are particularly pronounced and short blooming events (e. g. Ozsvárt et al., 2023) such as the Middle Triassic bloom (see Fig. 2). As the timespan of this event is so short and its occurrence values so high, the Middle Triassic blooming dictates much of the systematic shift p-value trends. We, therefore, analyzed the trend in radiolarian p-values before (0-200Ma) and after (327-529 Ma) the Middle Triassic bloom, and observed similar p-value trends and lower values than in the same test over 0-500 Ma (Fig. 2). This suggests that in addition to the periodic galactic forcing effect, an unidentified factor influenced radiolarian originations during periods when the other three groups had not yet existed.

Given the shared driving mechanism and shared marine environment, genera appearances between groups are not independent, thus we combined the four group *p*-values into a global *p*-value employing the Stouffer-method (Stouffer et al., 1949; Zaykin, 2011). The p-values emerge as follows: p=0.012 for dinoflagellate, p=0.014 for nannoplankton, p=0.002 for planktonic foraminifera, and p=0.237 for radiolaria (Fig. 5). When considering all genera data over the Phanerozoic and without applying shifts to the oscillation period, the global p-value emerges as $2.95 \times 10^{-5}$. Weighting this with the square root of the size of the datasets (N=48, 38, 48, 97, for the dinoflagellates, planktonic foraminifera, nannoplankton, and radiolaria, respectively), the global weighted p-value emerges as $p=1.25 \times 10^{-4}$. Over only the time period when all groups existed concurrently (0-174Ma), the global *p*-value is $3.76 \times 10^{-6}$. Since the data set size is the same in this case (N=38), the weighting does not affect this p-value. To account for testing nested data samples (with the full and the coexisting data samples), we applied a Bonferroni correction. The resulting global post-trial p-value is $7.6 \times 10^{-6}$, corresponding to the post-trial significance of 4.52σ, which exceeds the conventional 3σ discovery threshold used in astronomical contexts. Our results suggest that biodiversity on Earth has been affected by Galactic forcing, i.e., variations in the flux and energy of material reaching Earth over long timescales and not only via large isolated events.

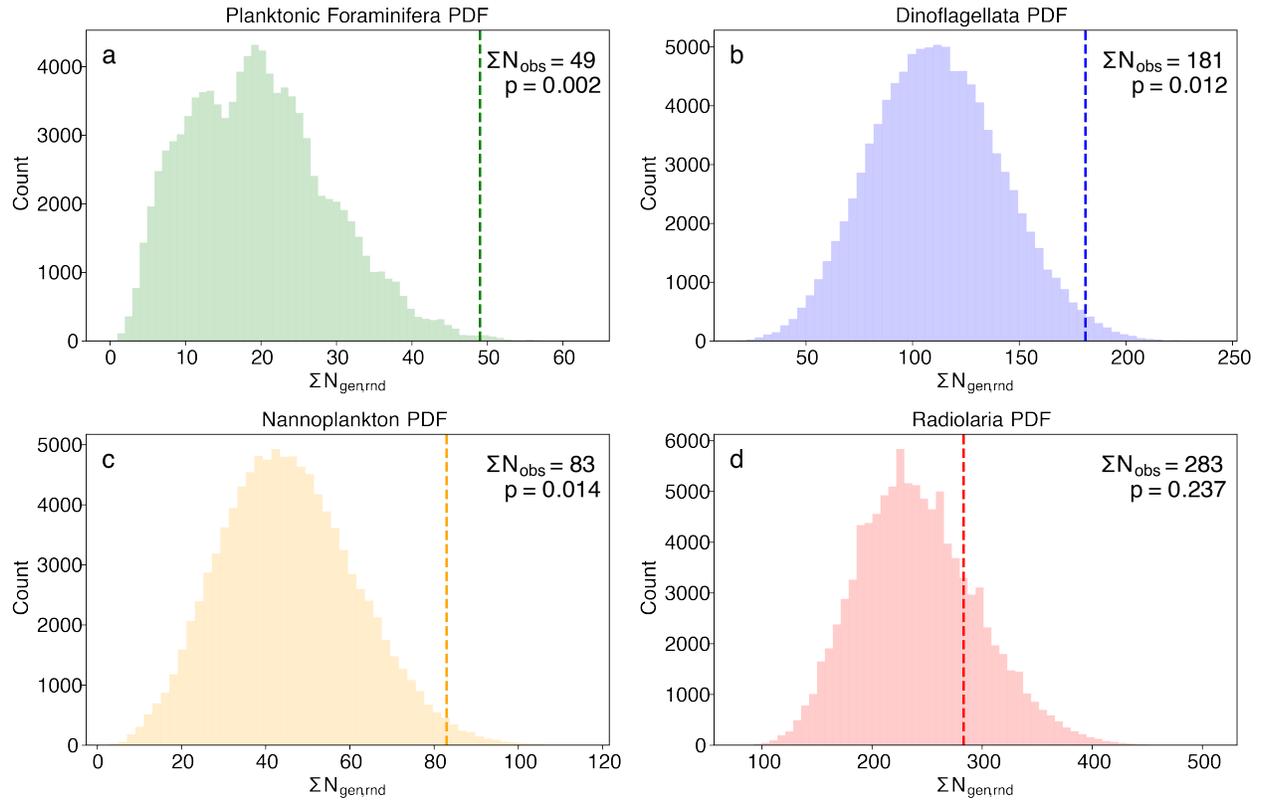

Figure 5. Discrete probability density functions (histograms) of the randomized total sum of new genera appearing in geological ages during maximum Solar System excursions from the Galactic plane are shown for Planktonic Foraminifera (a), Dinoflagellata (b), Nannoplankton (c), and Radiolaria (d), with observed values indicated by vertical dashed lines. The observed total sum of new genera and the corresponding p-values are displayed in the top right corner of each subplot.

## 5. Discussion: Cosmic ray induced mutations and bloomings

The most important effect of cosmic radiation on the biosphere is ionization (Usoskin & Kovaltsov, 2006) that leads to mutations. For example, solvated electrons generated by cosmic ray reactions with water damage the DNA of living organisms (Kumar et al., 2019). Solvated electrons introduce changes in the electronic structure of biomolecules and damage the DNA and proteins of microplankton organisms (see Fig. 6). Enhanced levels of muon and primary electron radiation from proton air showers can also introduce major effects on living organisms by creating a population of solvated electrons in the water. These solvated electrons are highly reactive and can participate in various chemical reactions, potentially leading to biological damage (e.g., Atri, 2016). The most common effect of ionization is DNA double-strand breaks (Fig. 6).

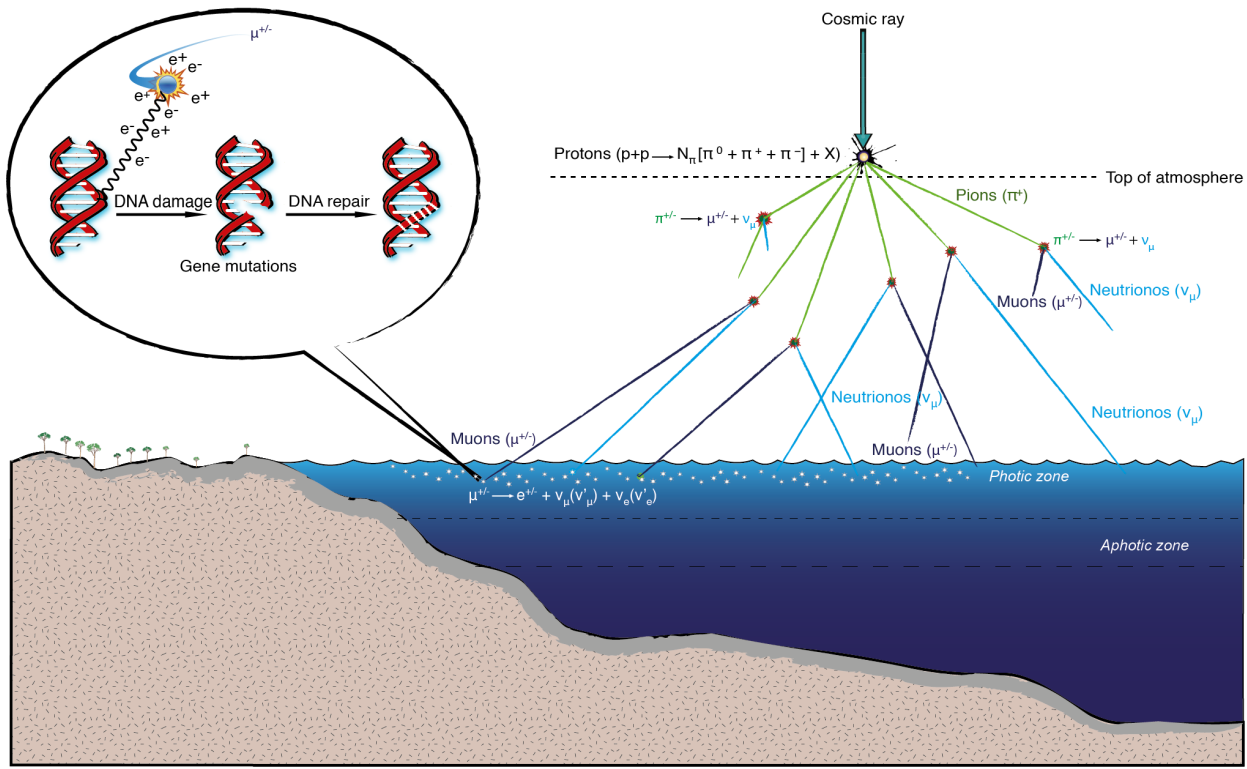

Figure 6. Visualization of some of the effects of cosmic rays on the DNA of microorganisms and a simplified theory of the process of the double-strand break DNA repair mechanism which causes frequent gene mutations.

There are various models in which cells can repair this type of damage, but as repairs are imperfect, mutations are introduced into the gene pool (Rodgers & McVey, 2016; Hanscom & McVey, 2020). If the high-energy cosmic ray flux is increased, this repair mechanism can become inadequate, thereby greatly increasing the rate of gene mutations entering the gene pool, even in deep-sea organisms (Hanscom & McVey, 2020). Furthermore, cosmic rays may trigger large chromosomal abnormalities and large-scale changes in chromosome structure that can affect the functioning of numerous genes, resulting in major phenotypic consequences (Lodish et al., 2000) and thus leading to genera diversification. Genome changes caused by cosmic rays also likely led to the abrupt changes in the external morphological characteristics of microorganisms that we see in each bloom period. In these morphological renewals, the actin-cytoskeleton system might regulate the repair process, which is a dynamic network made up of actin polymers and associated actin-binding proteins (e.g., Hurst et al., 2019). Exposing the biosphere to ionizing radiation can, therefore, plausibly lead to the mutation of simple living systems such as planktonic organisms (Atri & Melott, 2014). Although this change in cosmic radiation intensity is perhaps the most difficult to document in the sedimentary record, our findings indicate that it likely played an important role in the explosive diversification of dinoflagellates, nannoplankton, planktonic foraminifera, and radiolarians over much of the Phanerozoic.

Of the four groups we examined, radiolarians are the only record between 0-500 Ma. However, the correlation between their originations and solar system oscillations is significantly impacted by the extreme blooming event in the Middle Triassic. One plausible explanation for the magnitude of this event is an increased flux of cosmic rays during the Mesozoic geomagnetic dipole low (Denis et al., 2002). This event, characterized by a prolonged minimum in the magnetic dipole moment of Earth's magnetic field (see Fig. 3C), compounded with minimal shielding from the Galactic magnetic field, likely led to a more pronounced exposure of Earth's biosphere to cosmic radiation. Earth's magnetic field is stronger than the Galactic magnetic field but affects only a short part of the trip of the particles. Consequently, it is essential to understand how the interplay between the pole orientation and strength of the geomagnetic field over Earth's history and variations in Galactic magnetic shielding have modulated the cosmic ray flux reaching Earth's surface. Future examinations ought to consider other terrestrial and cosmic mechanisms that might affect the details of this picture and test additional simple forms of life.

## 6. Conclusions

The strength of the cosmic ray flux reaching Earth's surface is modulated by a multitude of factors, including the Solar system's location relative to the Galactic plane (and hence the strength of the Galactic magnetic field) and the strength of the Earth's and the Sun's magnetosphere. We detect a significant correlation between the oscillation of the Solar System from the Galactic plane and the origination of four different marine micro phyto- and zooplanktonic groups. We conducted statistical tests to uncover the significance between the appearance of the new genera in the four marine groups and the time of the Solar System maximum oscillations (see Materials and Methods). Considering the correlation between p-values due to the shared marine environment among these four groups, and the plausible effect of solvated electrons in water, the global weighted p-value emerges as $1.25 \times 10^{-4}$ when considering all genera data. When we restrict the analysis for those time periods during which all four groups co-existed, a post-trial significance of $4.52\sigma$ emerges. Therefore, far from the Galactic plane, where the magnetic shielding of the Galaxy is reduced, an enhanced flux of cosmic rays and their air showers are able to reach Earth's surface and induce mutations in the DNA of marine microorganisms, plausibly leading to the appearance of new genera/species (Fig. 6). The resulting increase in origination is controlled by natural selection, which has led to a slow evolutionary equilibration of microplankton organisms following the blooms. These patterns suggest that increasing the Galactic forcing of cosmic radiation intensity may significantly influence marine ecosystems and ultimately play an essential role in the formation of several significant microorganisms throughout the Phanerozoic. Such galactic forcing effects are not limited to marine microorganisms and are likely to have wide-reaching consequences in the diversification of other groups of fossil organisms.

## Acknowledgments

E. K., acknowledges support from the Alexander von Humboldt Foundation throughout the majority of this research, and acknowledges support also from the German Science Foundation DFG, via the Collaborative Research Center SFB1491: Cosmic Interacting Matters—from Source to Signal (Grant No. 445052434). Z.G.M. is supported by the Euretta J. Kellett Fellowship from Columbia University. O.P. was supported by the Hungarian National Research, Development and Innovation Office (project no.: K135309). We would like to thank our colleague, biologist András Tóth, for his detailed discussions during the early stages of the project, which were particularly inspiring for the continuation of our work. We are very grateful to anonymous reviewers for all important revision, suggestion and critical comments. The present study is HUN-REN-MTM-ELTE, Research Group for Paleontology contribution No. 426.

**Conflict of Interest Statement**

The authors have no conflicts of interest to disclose.

**Open Research**

### Microplankton databases

We obtained the first appearance datum for nannoplankton and planktonic foraminifera from the Microtax database, which is regularly updated, well-documented, and easily available:

### Nannoplankton

Mesozoic: https://www.mikrotax.org/Nannotax3/index.php?dir=ntax_mesozoic
Cenozoic: https://www.mikrotax.org/Nannotax3/index.php?dir=ntax_cenozoic

### Plankton foraminifera

Mesozoic: https://www.mikrotax.org/pforams/index.php?dir=pf_mesozoic
Cenozoic: https://www.mikrotax.org/pforams/index.php?dir=pf_cenozoic

### Radiolaria

The reconstructed temporal patterns of genera originations of radiolaria were obtained from the new catalogs of Paleozoic, Mesozoic and Cenozoic radiolarians (Caridroit et al., 2017; O'Dogherty et al., 2009; O'Dogherty, L. (Ed.), 2021):

Paleozoic:

http://sciencepress.mnhn.fr/sites/default/files/periodiques/documents-supplementaires/geo/g2017n3/geodiversitas-g2017n3a6-paleozoic-radiolarian-species-list.xls

Mesozoic:

https://sciencepress.mnhn.fr/en/periodiques/geodiversitas/31/2/online-supplement-1-rangechart-triassic-genera

https://sciencepress.mnhn.fr/en/periodiques/geodiversitas/31/2/online-supplement-2-rangechart-jurassic-cretaceous-genera

Cenozoic: https://sciencepress.mnhn.fr/sites/default/files/documents/en/geodiversitas2021v43a21-appendix_2_genera.zip

**Dinoflagellata**

The dinoflagellate data were obtained from the Lentin and Williams catalog (Fensome et al., 2019) and partly from the DINOSTRAT database (Bijl, 2022; Bijl, 2024) and from the PALSYS database (https://palsys.org/genus).